\documentclass[intlimits,twoside,a4paper]{article}

\usepackage{amsmath,amssymb,mathrsfs}
\usepackage{graphicx}
\usepackage{wrapfig}

\usepackage[T2A]{fontenc}
\usepackage[cp1251]{inputenc}

\usepackage{color}

\usepackage{cmpj2}

\issue{2013}{16}{3}{33801}
\doinumber{10.5488/CMP.16.33801}

\title[Merging diabolical points of a superconducting circuit]%
{Merging diabolical points of a superconducting circuit}
\author[R. Leone, A. Monjou]{R. Leone, A. Monjou}

\address{Statistical Physics Group, Institut Jean Lamour, UMR CNRS 7198, Universit\'e de Lorraine, \\  BP 70239, F-54506 Vand\oe uvre-l\`es-Nancy Cedex, France}

\date{Received July 26, 2012, in final form October 11, 2012}
\authorcopyright{R. Leone, A. Monjou, 2013}

\begin{document}

\maketitle

\begin{abstract}
We present the first theoretical study of the merging of diabolical points in the context of superconducting circuits. We begin by studying an analytically solvable four-level model which may serve as theoretical pattern for such a phenomenon. Then, we apply it to a circuit named Cooper pairs pump, whose diabolical points are already known.
\pacs 85.25.Cp, 74.50.+r, 74.78.Na, 03.67.-a
\end{abstract}

\section{Introduction}

Superconducting circuits~\cite{Devoret} become more and more central in modern quantum physics. Their principal building blocks are ultra-small Josephson junctions~\cite{Tinkham} which can be assembled in a variety of  ways, each of them defining a specific quantum system. Being frequently easier to manipulate, these solid state devices tend to supplant atomic and photonic systems, the ``old paragons'' of quantum mechanics. Notably, superconducting circuits are widely used to engineer qubits~\cite{Nakamura,Makhlin,Martinis,You,Wendin,Clarke}: the non-linear behavior of Josephson junctions serves to isolate couples of levels in a Hamiltonian spectrum. They are also used to perform the role of analogs of cavity quantum electrodynamics~\cite{Blais,Wallraff,Girvin} (a qubit plays the role of an artificial atom while a transmission line carries artificial photon modes), (non-) Abelian holonomies~\cite{Fazio,Cholascinski,Pekola2,Leek,Pekola}, (non-) Abelian quantum charge pumpings~\cite{Pump1,Pump2,Pump3,Pump4,Pump5,Pump6,Pump7,Pump8}, etc. In brief, they are good candidates for implementing quantum logic operations~\cite{Wendin,Nielsen} as well as appear to be quite promising for applications in electrical metrology~\cite{MetrologyCPP}. The Cooper pairs pump (CPP) considered in this article is an archetype of quantum circuit having a few (collective) degrees of freedom.
In reference~\cite{Pump6} there has been  theoretically demonstrated  a possible topological quantization of the pumped charge through an invariant called first Chern number (or Chern index)~\cite{Pump6,Goryo}. It relies on the existence of \emph{diabolical points}~\cite{Triangles} in the three-dimensional parameter space of a system, i.e., on double degeneracies characterized by a linear dispersion in all directions of that space.

Quite recently, G. Montambaux et al. have demonstrated the possibility of merging Dirac points in certain two-dimensional crystals, especially in hexagonal --- graphene-like lattices~\cite{Montambaux1,Montambaux2,Montambaux3,Montambaux4} (see also references~\cite{Optics,Pereira,Asano}). Dirac points are nothing else but diabolical points in the reciprocal space of crystals. They are ``naturally'' located at points of high symmetry, e.g., at vertices of a regular hexagonal lattice. However, in accordance with the famous Wigner-von Neumann theorem~\cite{WVN}, they may move, driven by well-chosen additional parameters. In the graphene example, the two triangular sublattices carry non-equivalent Dirac points. The merging of two neighboring non-equivalent Dirac points evokes the meeting of a knot and its anti-knot: being monitored by a \emph{merging parameter}, they move closer together, then merge into a single degeneracy and finally disappear. At the transition, the single degeneracy is characterized by a quadratic dispersion in the direction of merging. Inspired by the works of G. Montambaux et al.,  in this paper we present a theoretical study of the merging of diabolical points in the context of superconducting circuits. The choice of the CPP was motivated by its well-known diabolical points located in a hexagonal lattice, a property which confers to that circuit a great similarity to graphene.

The paper is structured as follows. Before introducing the CPP, we begin in section~\ref{sec:level2} by formally treating the merging process within the framework of a generic four-level model. The reason for such a choice in the organization is threefold: (i) the merging using the CPP relies on the model that permits to display \emph{a priori} the merging parameter of the CPP; (ii) it provides a ``universal Hamiltonian'' which is susceptible to be realized in different quantum contexts of ours; (iii) it gives the opportunity to briefly review some characteristics of double degeneracies in a parameter space. In section~\ref{sec:level3}, we present the CPP and emphasize the symmetry origin of its ``mobile diabolical points''. Finally, via a modification of the circuit, we suggest in section~\ref{sec:level4} a way of merging these points. This will be done through an effective Josephson energy as the merging parameter.

\section{\label{sec:level2}The four-level model}

We consider a model Hamiltonian depending on a triple of tunable parameters ${\bf R}=(X,Y,Z)$ and having the form
\begin{align}
H({\bf R})=\left(\begin{array}{cccc}\xi+X&F\re^{\ri Z}&F\re^{\ri Z}&0\\F\re^{-\ri Z}&-\xi+Y&G\re^{\ri Z}&F\re^{\ri Z}\\F\re^{-\ri Z}&G\re^{-\ri Z}&-\xi-Y&F\re^{\ri Z}\\0&F\re^{-\ri Z}&F\re^{-\ri Z}&\xi-X\end{array}\right)\label{H4}
\end{align}
in an orthonormal basis $\{|e_1\rangle,|e_2\rangle,|e_3\rangle,|e_4\rangle\}$. Here, $F\neq0$ and $\xi$ are constants, and $G$ is an additional tunable parameter. The latter is a dubbed \emph{merging parameter} for the reason which will appear shortly. We will restrict $Z$ to the interval $[-\frac{\pi}{2};\frac{\pi}{2}]$ since the translation $Z\to Z+\pi$ amounts to the change of  the sign of $F$ and $G$. The set of vectors ${\bf R}$ forms the natural parameter space of the problem. In this space, the spectrum of $H$ possesses the symmetry $\mathcal D_{2h}$. Indeed, $H$ is (anti)unitary transformed under sign-reversing of $X$, $Y$ and $Z$. Explicitly, we have
\begin{enumerate}
\item $H(X,Y,-Z)=\mathcal K\,H(X,Y,Z)\,\mathcal K^\dag$, where $\mathcal K$ is the complex conjugation operator with respect to the basis $\{|e_1\rangle,|e_2\rangle,|e_3\rangle,|e_4\rangle\}$;
\item $H(X,-Y,Z)=\big[\mathcal U(Z)\mathcal K\big]H(X,Y,Z)\big[\mathcal U(Z)\mathcal K\big]^\dag$ with
\begin{align*}
\mathcal U(Z)=\left(\begin{array}{cccc}\re^{2\ri Z}&0&0&0\\0&0&1&0\\0&1&0&0\\0&0&0&\re^{-2\ri Z}\end{array}\right);
\end{align*}
\item $H(-X,Y,Z)=\!\big[\mathcal U(Z)\mathcal T\big]H(X,Y,Z)\big[\mathcal U(Z)\mathcal T\big]^\dag$ with
\begin{align*}
\mathcal T=\left(\begin{array}{cccc}0&0&0&1\\0&0&1&0\\0&1&0&0\\1&0&0&0\end{array}\right).
\end{align*}
\end{enumerate}
In particular, under ${\bf R}$-inversion, we observe the simple unitary equivalence $H(-{\bf R})=\mathcal TH({\bf R})\mathcal T^\dag$.

Since $\langle e_1|(H-\mu)(H-\nu)|e_4\rangle=2\,F^2\re^{2\ri Z}\neq0$, for any $\mu$ and $\nu$, $H$ has at least three distinct eigenvalues. Thus, $\lambda$ is a (doubly) degenerate eigenvalue of $H$ if and only if (iff) there exists a real $\beta>0$ such that
\begin{align}
(H-\lambda)(H+\lambda+\beta)(H+\lambda-\beta)=0.\label{poly}
\end{align}
In this case, $-\lambda-\beta$ and $-\lambda+\beta$ are the other eigenvalues and $\lambda$ is the smallest one iff $\beta<-2\lambda$. After a little algebra based on equation~(\ref{poly}) and $H$'s characteristic polynomial, we find that the ground level of $H$ is degenerate iff $G$ is greater than the critical value $G_{\rm c}=\sqrt{2F^2+\xi^2}-\xi$ while $X=\pm X_{\rm d}$ with
\begin{align*}
X_{\rm d}=\sqrt{\left(1-\frac{G_{\rm c}}{G}\right)(G+2\xi)(G+G_{\rm c}+2\xi)}\,.
\end{align*}

\begin{figure}[!t]
\centerline{
\includegraphics[width=0.95\textwidth]{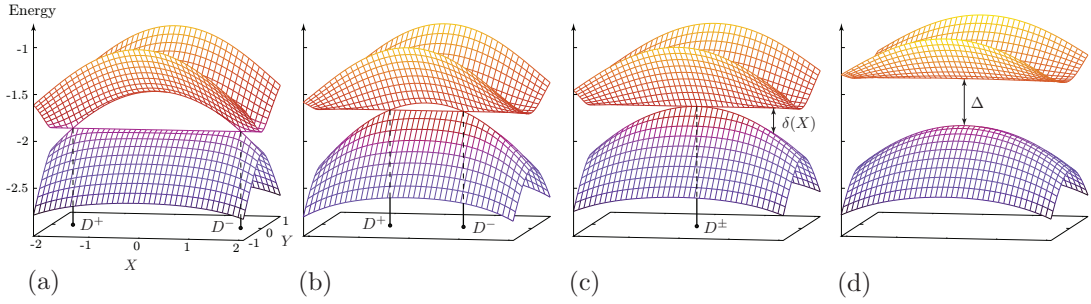}
}
\caption{(Color online) Plots of the two lowest energy levels as functions of $X$ and $Y$, with $Z=0$ and $\xi=|F|$, for different values of the merging parameter $G$. As long as $G>G_{\rm c}$, the distance between the diabolical points $D^+$ and $D^-$ decreases with $G$. They merge for $G=G_{\rm c}$ into a single degeneracy characterized by a quadratic dispersion $\delta(X)$ in the $X$-direction and disappear as $G<G_{\rm c}$. The plot unit is $|F|$. (a) $G=G_{\rm c}+0.25\,|F|$; (b) $G=G_{\rm c}+0.05\,|F|$; (c) $G=G_{\rm c}$; (d) $G=G_{\rm c}-0.3\,|F|$.}\label{merginginline}
\end{figure}

As long as $G>G_{\rm c}$, the two distinct points $D^\pm$, located at ${\bf R}^\pm=(\mp X_{\rm d},0,0)$, are isolated \emph{degenerate points} in the ${\bf R}$-space. They illustrate a classical theorem of von Neumann and Wigner~\cite{WVN} which states that, generically, twofold degeneracies have codimension three. Alternatively stated, in an $N$-dimensional parameter space, there generically exist submanifolds of dimension $N-3$ over which a level is doubly degenerate. Here, $N=3$ and the degenerate subspaces are the points (dimension: $3-3=0$). One should also think of a bigger space collecting all the parameters entering $H$, as the six-dimensional space of vectors $\mathscr R=(X,Y,Z,\xi,F,G)$. The submanifold verifying, simultaneously $F\neq0$, $Y=Z=0$, $G\geqslant G_{\rm c}(\xi,F)$, and $X=\pm X_{\rm d}(\xi,F,G)$ carries a degenerate ground level and has a (co)dimension 3 in the $\mathscr R$-space, as expected.

The points $D^\pm$ move closer together when we reduce $G$, while conserving the symmetry relations $[H,\mathcal U]=[H,\mathcal K]=0$ (see figure~\ref{merginginline}). This corresponds to a generic situation: if they deviated from the planes $Y=0$ or $Z=0$, each of them would split into 2 (or 4) distinct degenerate points. They merge at ${\bf R}=\bf 0$ for $G=G_{\rm c}$ and finally disappear as soon as $G<G_{\rm c}$, the minimal gap between the two lowest levels being
\begin{align*}
\Delta=\frac12\left(\sqrt{16F^2+(G-2\xi)^2}-3G\right)-\xi.
\end{align*}

Let $\Pi$ be the projector into $H({\bf R}^+)$'s ground eigenspace and $\{|1\rangle,|2\rangle\}$ an orthonormal basis of that subspace. Let  $\pmb{\sigma}=(\sigma_x,\sigma_y,\sigma_z)$ be the triple of operators represented by the usual Pauli matrices in the basis. Up to an unimportant component along $\Pi$, there exists a unique fixed real matrix $\textsf M$ of the order of 3 such that $\Pi\big[\pmb\nabla H({\bf R}^+)\cdot({\bf R}-{\bf R}^+)\big]\Pi$ reads $\pmb\sigma\cdot{\textsf M}({\bf R}-{\bf R}^+)$. Obviously, ${\textsf M}$ depends on the choice of the basis, but it is a simple task to show that the signum of its determinant is intrinsic to the degeneracy. It is the \emph{signature}~\cite{Simon} of the degenerate point $D^+$ in the ${\bf R}$-space.
One can explicitly choose
\begin{align*}
|1\rangle&=\frac{1}{\sqrt{2}}\Big(|e_2\rangle-|e_3\rangle\Big), \\
|2\rangle&=\sqrt{\frac{G(G+2\xi)}{4\lambda^2-\beta^2}}
\left[\left(1-\frac{X}{G+2\xi}\right)|e_1\rangle-\frac{F}{G}|e_2\rangle
-\frac{F}{G}|e_3\rangle+\left(1-\frac{X}{G+2\xi}\right)|e_4\rangle\right],
\end{align*}
where $\beta$ is given by
\begin{align*}
\beta=\sqrt{G_{\rm c}^2+2\xi\left(1-\frac{G_{\rm c}}{G}\right)(G_{\rm c}+2\xi)}\,.
\end{align*}
Within this choice, one finds
\begin{align*}
\det{\textsf M}=\frac{4F^2(G+2\xi)X_{\rm d}}{(4\lambda^2-\beta^2)^2}\,.
\end{align*}
As long as $G>G_{\rm c}$, the signature of $D^+$ is $+1$. In its vicinity, the two lowest levels are close together and $\pmb\sigma\cdot{\textsf M}({\bf R}-{\bf R}^+)$ is an accurate Hamiltonian for the states belonging to them. Since $D^+$ has a nonzero signature, the level splitting around it is effective from the first order in $\|{\bf R}-{\bf R}^+\|$ in all directions of the parameter space: $D^+$ is a \emph{diabolical point}~\cite{Triangles}. By symmetry, so does $D^-$, whose signature is found to be $-1$. For $G=G_{\rm c}$, the single degeneracy located at the origin has a vanishing signature, because the dispersion in the merging direction is quadratic. A perturbative analysis shows that a deviation $(0,0,0)\to(X,0,0)$ opens a gap
\begin{align*}
\delta(X)=\frac{G_{\rm c}^3X^2}{2F^2(F^2+G_{\rm c}^2)}+{\rm O}(X^4).
\end{align*}

Before introducing the system which will serve to realize our four-level model, let us end this section with two remarks. The first one is peculiar to the model: if $Y$, $Z$ are suppressed and $\xi$, $F$ tunable, the Hamiltonian may be used to construct non-Abelian holonomies~\cite{Recati,LeonePumping} over the manifold satisfying simultaneously $F\neq0$, $G\geqslant G_{\rm c}(\xi,F)$ and $X=\pm X_{\rm d}(\xi,F,G)$. The model may also serve to implement non-Abelian pumpings having $Z$ as pumping parameter~\cite{LeonePumping}. The second remark is more general and concerns the signature. Consider some Hamiltonian $H$ continuously defined over the ${\bf R}$-space. Suppose the existence of a nonsingular transformation $\mathfrak t:{\bf R}\mapsto{\bf R}'$ associated with a fixed symmetry operator $T$, such that $H({\bf R}')=TH({\bf R})T^\dag$. If $T$ is unitary, it is straightforward to verify that the signatures are conserved by the transformation if $\mathfrak t$ is orientation-preserving and reversed otherwise. If $T$ is antiunitary, the contrary occurs. In our example, the signature of $D^-$ is due to the orientation-reversing map ${\bf R}\mapsto-{\bf R}$ associated with the unitary operator $\mathcal T$. Moreover, below we will  use successive orientation-preserving transformations of the parameter space, without incidence on the signature.

\section{\label{sec:level3}The Cooper pairs pump and its diabolical points}

\subsection{Basic settings}

\begin{figure}[!b]
\centerline{
\includegraphics[width=0.5\textwidth]{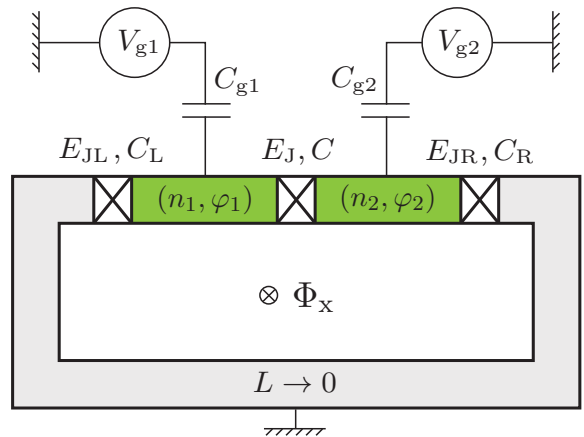}
}
\caption{(Color online) The Cooper pairs pump (CPP) is essentially an array of three Josephson junctions in a loop configuration. It depends on three external parameters: the gate voltages $V_{{\rm g}k}$ on the superconducting islands and the magnetic flux $\Phi_{\rm x}$ threading the loop. The system is said to be mirror symmetric if the ``exterior junctions'' are identical (i.e., $E_{\rm JL}=E_{\rm JR}$, $C_{\rm L}=C_{\rm R}$) and totally symmetric if all the junctions are identical.}
\label{figure2}
\end{figure}

We consider one of the simplest implementations for a CPP, represented in figure~\ref{figure2}. It is a small-inductance superconducting loop ($L\to0$), threaded by a magnetic flux $\Phi_{\rm x}$ and broken by three ultra-small Josephson junctions~\cite{Tinkham}. The junctions are assumed non-dissipative. They enclose two superconducting islands, polarized by gate voltages $V_{{\rm g}k}$ through low gate capacitances $C_{{\rm g}k}$ ($k=1,2$). We set $n_{{\rm g}k}={C_{{\rm g}k}V_{{\rm g}k}}/{2e}$ the corresponding gate charges in the unit of $2e$ ($>0$). The vanishing loop inductance leads to a biasing phase~\cite{Tinkham} $\varphi_{\rm x}=2\pi{\Phi_{\rm x}}/{\Phi_0}$ across the CPP, where $\Phi_0={h}/{2e}$ is the quantum of flux. We suppose $\Phi_{\rm x}$, $V_{\rm g1}$ and $V_{\rm g2}$ independently tunable. The system has two collective degrees of freedom~--- one for each island~--- and depends on three external parameters that we choose to be $n_{\rm g1}$, $n_{\rm g2}$ and $\varphi_{\rm x}$ rather than $V_{\rm g1}$, $V_{\rm g2}$ and $\Phi_{\rm x}$. The conjugated operators assigned to the degrees of freedom are $ n_k$ and $\varphi_k$ $(k=1,2)$: the number of Cooper pairs in excess (with respect to charge neutrality) and the phase of the superconducting parameter of the $k^{\rm th}$ island, respectively. They verify the commutation relations $[ n_j,\varphi_k]=\ri\,\delta_{j,k}$. We will study the CPP in charge representation; $|n_1,n_2\rangle$ will represent the fundamental charge states defined such that
$\re^{\pm\ri\varphi_1}|n_1,n_2\rangle=|n_1\mp 1,n_2\rangle$ and $\re^{\pm\ri\varphi_2}|n_1,n_2\rangle=|n_1,n_2\mp 1\rangle$. Since eigenvalues of $n_1$ and $n_2$ can theoretically cover all the range of $\mathbb Z$, the orthonormal basis $\mathscr B=\{|n_1,n_2\rangle\,|\,(n_1,n_2)\in\mathbb Z^2\}$ spans the whole Hilbert space of the problem.

Throughout this article, $C_{\Sigma}=C_{\rm L}+C+C_{\rm R}$ will be the capacitance unit and $E_{\rm C}={(2e)^2}/{C_\Sigma}$ will be the energy unit. The latter is a typical charging energy of the circuit. We will study the system in the Coulomb blockade regime, characterized by Josephson energies small in comparison to $E_{\rm C}=1$. Using the canonical quantization procedure, a Hamiltonian $H=H(n_{\rm g 1},n_{\rm g2},\varphi_{\rm x})$ may be derived for the system. It splits into two parts: a charging Hamiltonian $H_{\rm C}=H_{\rm C}(n_{\rm g1},n_{\rm g2})$ and a Josephson tunneling Hamiltonian $H_{\rm J}=H_{\rm J}(\varphi_{\rm x})$. Neglecting the gate capacitances in comparison to $C_{\Sigma}=1$ and using the notation $\pmb\alpha=(\alpha_1,\alpha_2)$, the former is
\begin{align}
H_{\rm C}=\frac12\,({\bf n}-{\bf n}_{\rm g})\cdot{\textsf C}^{-1}({\bf n}-{\bf n}_{\rm g}),\label{HC}
\end{align}
where ${\textsf C}$ is the capacitance matrix:
\begin{align*}
{\textsf C}=\left(\begin{array}{cc}C_{\rm L}+C&-C\\-C&C+C_{\rm R}\end{array}\right).
\end{align*}
The charging Hamiltonian is obviously diagonal in the basis $\mathscr B$ and verifies $H_{\rm C}({\bf n}_{\rm g}+{\bf a})=\re^{-\ri{\bf a}\cdot\pmb\varphi}H_{\rm C}({\bf n}_{\rm g})\,\re^{\ri{\bf a}\cdot\pmb\varphi}$ for any integer vector ${\bf a}$. Over the ${\bf n}_{\rm g}$-plane, the energy surface of the eigenstate $|{\bf 0}\rangle=|0,0\rangle$ is an elliptic paraboloid centered at ${\bf n}_{\rm g}={\bf 0}$. Thus, the energy surface of $|{\bf n}\rangle=|n_1,n_2\rangle=\re^{-\ri{\bf n}\cdot\pmb\varphi}|{\bf 0}\rangle$ is simply the translation by ${\bf n}$ of this paraboloid. Two different states $|{\bf n}\rangle$ and $|{\bf n}'\rangle$ are degenerate on a straight line characterized by ${\bf n}$, ${\bf n}'$ and the capacitances. Then, one easily checks that $|n_1,n_2\rangle$ is the ground state of $H_{\rm C}$ in a hexagon ${\textsf{hex}}(n_1,n_2)$ centered at ${\bf n}_{\rm g}={\bf n}$. This defines the well-known honeycomb lattice of the CPP. It is graphically obtained by integer translations of two nonequivalent lattice points $T^\pm$ whose coordinates are
\begin{eqnarray*}
{\bf n}_{\rm g}(T^\pm)=\pm\frac{1}{2}\;{\textsf{C}}\left(\begin{array}{c}(\textsf C^{-1})_{11}\\(\textsf C^{-1})_{22}\end{array}\right).
\end{eqnarray*}

The lattice picture is useful if we identify each fundamental state $|n_1,n_2\rangle$ with its corresponding hexagon ${\textsf{hex}}(n_1,n_2)$. With respect to $H_{\rm C}$, the common side of two neighboring hexagons is a piece of the degeneracy line between the states, while the vertices are points of triple degeneracy. Introducing the distance induced by the scalar product $({\bf x}|{\bf y})=2^{-1/2}\,{\bf x}\cdot{\textsf C}^{-1}{\bf y}$ in the plane, this picture allows one to interpret the charging energy of $|n_1,n_2\rangle$ as the squared distance between ${\bf n}_{\rm g}$ and the center of ${\textsf{hex}}(n_1,n_2)$. ``Branching'' $H_{\rm J}$, which can be brought into the form
\begin{align}
H_{\rm J}=U(\varphi_{\rm x})\Big[&-E_{\rm JL}\cos(\varphi_1+\varphi_{\rm x})-E_{\rm J}\cos(\varphi_2-\varphi_1+\varphi_{\rm x})-E_{\rm JR}\cos(\varphi_2+\varphi_{\rm x})\Big]\,U(\varphi_{\rm x})^\dag,\label{HJ}
\end{align}
couples the neighboring states and  \emph{generically} lifts the degeneracies of $H_{\rm C}$. Explicitly, $U(\varphi_{\rm x})=\re^{\ri{\pmb\kappa}\cdot{\bf n}\varphi_{\rm x}}$, with $\kappa_1=1-C_{\rm R}({\textsf C}^{-1})_{12}$ and $\kappa_2=2-C_{\rm R}({\textsf C}^{-1})_{22}$. In the Coulomb blockade regime, $H_{\rm J}$ is seen as a perturbation of $H_{\rm C}$. As a good approximation, the Hilbert space may be reduced to its subspace spanned by a few number of fundamental states in the neighborhood of ${\bf n}_{\rm g}$. To this end, we only take into account the states $|n_1,n_2\rangle$ at a distance of ${\bf n}_{\rm g}$ shorter than a certain value.

Since $H({\bf n}_{\rm g}+{\bf a},\varphi_{\rm x})=\re^{-\ri{\bf a}\cdot\pmb\varphi}H({\bf n}_{\rm g},\varphi_{\rm x})\,\re^{\ri{\bf a}\cdot\pmb\varphi}$, translations of lattice vectors ${\bf a}$ leave the physics unchanged up to a displacement $|{\bf n}\rangle\to|{\bf n}+{\bf a}\rangle$ of the fundamental charge states. Moreover, performing the gauge transformation $|n_1,n_2\rangle\to U(\varphi_{\rm x})|n_1,n_2\rangle$, the Hamiltonian is invariant under the translations $\varphi_{\rm x}\to\varphi_{\rm x}+2k\pi$ ($k\in\mathbb Z$). Thus, the spectrum of $H$ possesses the translational symmetry of a hexagonal prism lattice in the space of vectors ${\bf r}=({\bf n}_{\rm g},\varphi_{\rm x})$. In the new representation, let us introduce the complex conjugation operator $\mathcal K$, the ``sign change operator'' $\mathcal S:|n_1,n_2\rangle\mapsto|-n_1,-n_2\rangle$ and the ``charge exchange operator'' $\mathcal P:|n_1,n_2\rangle\mapsto|n_2,n_1\rangle$. Taking the $\varphi_{\rm x}$-axis vertical, the spectrum of $H$ possesses the point symmetry $\mathcal C_{2h}$: under the reflection $\sigma_h$ and the inversion $\iota$, we have $H({\bf r})=\mathcal K^\dag H(\sigma_h{\bf r})\mathcal K=\mathcal S^\dag H(\iota{\bf r})\mathcal S$. In particular, the symmetry $\mathcal C_2=\iota\circ\sigma_h$ implies that the Hamiltonians at $(T^+,\varphi_{\rm x})$ and $(T^-,\varphi_{\rm x})$ are antiunitary equivalents, and even unitary equivalents iff $\varphi_{\rm x}=0\mod\pi$.

\begin{figure}[!t]
\centerline{
\includegraphics[width=0.45\textwidth]{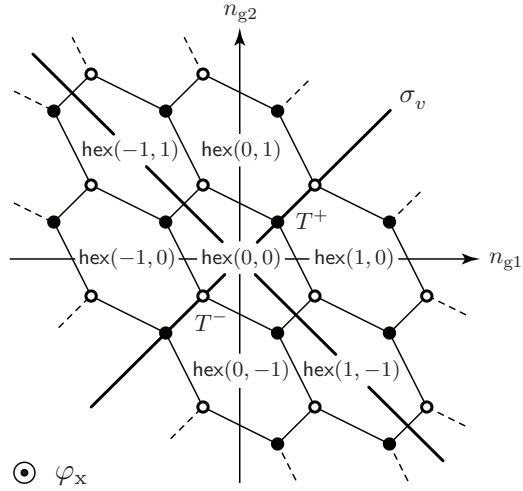}
}
\caption{The honeycomb lattice of the CPP over the ${\bf n}_{\rm g}$-plane. It is generated by the triple points $T^\pm$ whose coordinates are given in the text. Each state $|n_1,n_2\rangle$ is the ground eigenstate of $H_{\rm C}$ in the hexagon ${\textsf{hex}}(n_1,n_2)$. In the space of vectors ${\bf r}=({\bf n}_{\rm g},\varphi_{\rm x})$, the spectrum of the Hamiltonian $H$ possesses the symmetry $\mathcal C_{2h}$. It becomes $\mathcal D_{2h}=\mathcal C_{2h}\times\sigma_v$ if the CPP is mirror symmetric (the vertical reflection planes are represented by thick lines).}\label{hexagonesng}
\end{figure}

If the CPP is mirror symmetric, that is to say, if the ``exterior junctions'' are identical ($E_{\rm JL}=E_{\rm JR}$ and $C_{\rm L}=C_{\rm R}$), the symmetry $\mathcal D_{2h}$ is reached. Indeed, the reflection $\sigma_v$, shown in figure~\ref{hexagonesng}, exchanges $n_{\rm g1}$ and $n_{\rm g2}$, inducing the transformation $H(\sigma_v{\bf r})=[\mathcal V(\varphi_{\rm x})\mathcal P\mathcal K]H({\bf r})[\mathcal V(\varphi_{\rm x})\mathcal P\mathcal K]^\dag$, where $\mathcal V(\varphi_{\rm x})=\re^{-2\ri(n_1+n_2)\varphi_{\rm x}}$.

\subsection{The diabolical points}

\begin{figure}
\centerline{
\includegraphics[width=0.9\textwidth]{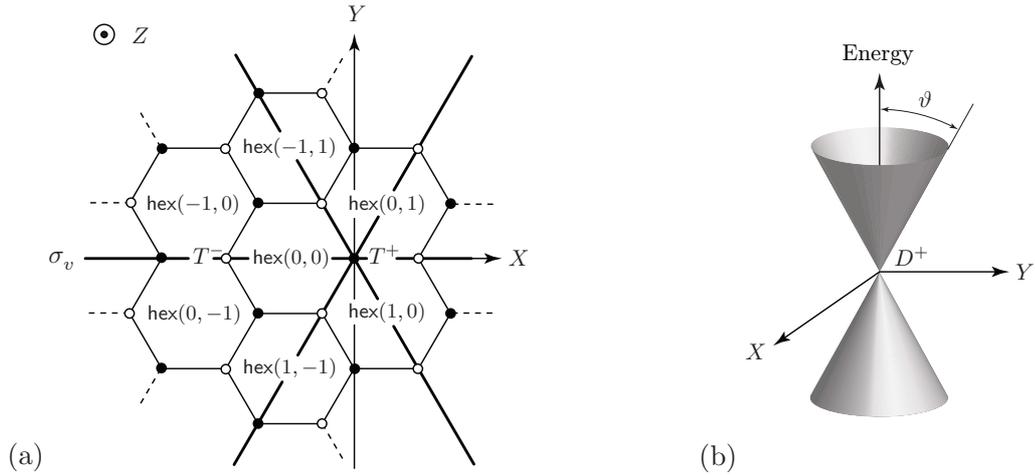}
}
\caption{(a) The regularized honeycomb lattice of the totally symmetric CPP. The new parameters $X$ and $Y$ are notably chosen so that $T^+$ is located at the origin. (b) Plot of the two lowest energy levels as a function of $X$ and $Y$ at $\varphi_{\rm x}=\pi$ in the close vicinity of $D^+$. In a totally symmetric configuration, they  locally form a right circular double cone whose aperture is $2\vartheta={2\pi}/{3}+{\rm O}(E_{\rm J})$.}\label{hexagonesXY}
\end{figure}

The CPP is said to be totally symmetric if the three junctions are identical. In this specific case, one has ${\bf n}_{\rm g}(T^\pm)=\pm(\frac13,\frac13)$. As shown in figure~\ref{hexagonesXY}(a), the orientation-preserving map $(n_{\rm g1},n_{\rm g2})\mapsto(X,Y)$, such that $X=\frac{\sqrt3}{2}(n_{\rm g2}+n_{\rm g1}-\frac23)$ and $Y=\frac12(n_{\rm g2}-n_{\rm g1})$, makes the hexagons regular in the $(X,Y)$-plane and places the origin at $T^+$. The charging energy of a state $|n_1,n_2\rangle$ becomes the usual squared distance between $(X,Y)$ and the center of ${\textsf{hex}}(n_1,n_2)$.

Setting $Z=\varphi_{\rm x}$, the spectrum possesses the symmetry $\mathcal D_{3h}$ in the so-defined ${\bf R}$-space [see figure~\ref{hexagonesXY}(a)]. The rotation $\mathcal C_3$ induces the transformation $H(\mathcal C_3{\bf R})=[\mathcal V(\varphi_{\rm x})\mathcal R]H({\bf R})[\mathcal V(\varphi_{\rm x})\mathcal R]^\dag$, with $\mathcal R:|n_1,n_2\rangle\mapsto|1-n_1-n_2,n_1\rangle$. The two symmetry operators $\mathcal P$ and $\mathcal R$ generate an unitary representation $\Gamma$ of the group $\mathcal D_3$ in the Hilbert space, such that $\Gamma(\sigma_v)=\mathcal P$ and $\Gamma(\mathcal C_3)=\mathcal R$. At the high symmetry points $(T^+,\varphi_{\rm x}=0\mod\pi)$, the Hamiltonian commutes with $\Gamma$.

The ground eigenspace of $H_{\rm C}(T^+)$ --- spanned by $|0,0\rangle$, $|1,0\rangle$, and $|0,1\rangle$ --- is an invariant subspace of $\Gamma$. If $\Gamma_{\rm g}$ is the restriction of $\Gamma$ to this space, we have
\begin{align*}
\Gamma_{\rm g}(\sigma_v)=\left(\begin{array}{ccc}1&0&0\\0&0&1\\0&1&0\end{array}\right)
\qquad{\text{and}}\qquad
\Gamma_{\rm g}(\mathcal C_3)=\left(\begin{array}{ccc}0&0&1\\1&0&0\\0&1&0\end{array}\right)
\end{align*}
in the basis $\{|0,0\rangle,|1,0\rangle,|0,1\rangle\}$. The subrepresentation $\Gamma_{\rm g}$ decomposes as $A\oplus E$, where $A$ and $E$ are respectively the totally  symmetric and the two-dimensional irreducible representations of $\mathcal D_3$. Obviously, the state
\begin{align*}
|A\rangle=\frac{1}{\sqrt3}\Big(|0,0\rangle+|1,0\rangle+|0,1\rangle\Big)
\end{align*}
belongs to $A$. Then, we complete the basis of $\Gamma_{\rm g}$ by choosing two orthonormal states belonging to $E$:
\begin{align*}
|E_1\rangle=\frac{1}{\sqrt2}\Big(|0,1\rangle-|1,0\rangle\Big)
\qquad{\text{and}}\qquad
|E_2\rangle=\frac{1}{\sqrt6}\Big(2|0,0\rangle-|0,1\rangle-|1,0\rangle\Big).
\end{align*}
In the basis $\{|A\rangle,|E_1\rangle,|E_2\rangle\}$, we thus have $\Gamma_{\rm g}=A\oplus E$ with $A(\sigma_v)=A(\mathcal C_3)=(1)$ and
\begin{align}
E(\sigma_v)=\left(\begin{array}{cc}-1&0\\0&1\end{array}\right), \qquad
E(\mathcal C_3)=\frac12\left(\begin{array}{cc}-1&-\sqrt3\\\sqrt3&-1\end{array}\right).\label{matrices}
\end{align}
Any Josephson coupling between the states $|A\rangle$, $|E_1\rangle$ and $|E_2\rangle$ is forbidden at the high symmetry points. Since $\langle E_\alpha|H_{\rm J}(\pi)|E_\alpha\rangle-\langle A|H_{\rm J}(\pi)|A\rangle=-3/2<0$ $(\alpha=1,2)$, up to the first order in $E_{\rm J}$, the ground level of $H(T^+,\pi)$ belongs to $E$ while the first excited one belongs to $A$. The contrary occurs for $H(T^+,0)$. Thus, the half-fluxoid condition $\Phi_{\rm x}={\Phi_0}/{2}\mod\Phi_0$ ensures the double degeneracy of the ground level at the point $T^+$. The same conclusion holds at $T^-$ from the equivalence between $H(T^-,\varphi_{\rm x})$ and $H(T^+,\varphi_{\rm x})$.

\begin{table}[!b]
\caption{The character table of $\mathcal D_3$. There are three irreducible representations: $A$ (totally symmetric), $B$ (antisymmetric) and $E$ (two-dimensional).}\label{character}
\vspace{1ex}
\begin{center}
\begin{tabular}{|c|ccc|}
\hline
$\mathcal D_3$&$E$&$2\,\mathcal C_3$&$3\,\sigma_v$\\
\hline\hline$A$&1&1&1\\$B$&1&1&-1\\$E$&2&-1&0\\
\hline
\end{tabular}
\end{center}
\end{table}

Let us analyze the signatures of the degenerate points $D^\pm=(T^\pm,\pi)$. Redefining, for convenience, $Z$ as $\varphi_{\rm x}-\pi$, $D^+$ is located at the origin of the new ${\bf R}$-space. Using the same notations as in section~\ref{sec:level2}, basis states $|1\rangle$ and $|2\rangle$ of $H(D^+)$'s ground level are partners of the irreducible representation $E=\Pi\,\Gamma\,\Pi$. They may~--- and they will~--- be chosen so that the matrices of $E$ are given by~(\ref{matrices}) in the basis $\{|1\rangle,|2\rangle\}$. Thereby, we have $|\alpha\rangle=|E_\alpha\rangle+{\rm O}(E_{\rm J})$, $\alpha=1,2$. The (anti)unitary transformations of $H$ under the action of $\mathcal D_{3h}$ imply the existence of two reals $\alpha$ and $\beta$ such that
\begin{align}
\Pi[\pmb\nabla H(D^+)\cdot{\bf R}]\Pi=\left(\begin{array}{cc}-\alpha X&\alpha Y+\ri\beta Z\\\alpha Y-\ri\beta Z&\alpha X\end{array}\right),\label{alphabeta}
\end{align}
in the basis $\{|1\rangle,|2\rangle\}$. The coefficients $\alpha$ and $\beta$ may be calculated as $\alpha=3^{-1/2}+{\rm O}(E_{\rm J})$ and $\beta=3^{-1/2}E_{\rm J}+{\rm O}(E_{\rm J}^2)$. Fixing $Z=0$, the two lowest levels, plotted around $T^+$ as functions of $X$ and $Y$, locally form  a right circular double cone whose aperture is $2\,{\mathrm{arccot}}(\alpha)={2\pi}/{3}+{\rm O}(E_{\rm J})$ [see figure~\ref{hexagonesXY}~(b)]. This is a consequence of the $\mathcal D_3$ symmetry.  Writing the right-hand side of equation~(\ref{alphabeta}) in the form $\pmb\sigma\cdot{\textsf M}{\bf R}$, one has $\det{\textsf M}=\alpha^2\beta>0$: the signature of $D^+$ is $+1$. Returning to the natural ${\bf r}$-space, the points $D^\pm$ are located at ${\bf r}^\pm=({\bf n}_{\rm g}(T^\pm),\pi)$. Since ${\bf r}^-=\mathcal C_2{\bf r}^+$ and $H(\mathcal C_2{\bf r})=[\mathcal S\mathcal K]H({\bf r}^+)[\mathcal S\mathcal K]^\dag$, the signature of $D^-$ is $-1$.

The existence of ``signed degeneracies'' is fundamental to the physics of quantum pumpings. They quantize the pumped charge along classes of cycles in the parameter space~\cite{Pump6}. They are robust in the sense that their existence is ensured by the Wigner-von Neumann theorem even though the $\mathcal D_3$ symmetry is broken. In some ways, one should say that the symmetry plays an important role of producing signed degeneracies which become ``accidental'' as soon as the symmetry is broken. Under continuous variations of the circuit characteristics (capacitances and Josephson energies) they  continuously move in the plane $\varphi_{\rm x}=\pi$, conserving their signature and the relation $[H,\mathcal K]=0$ (though loosing the regularity of the conical intersection over the $(X,Y)$-plane).

\section{\label{sec:level4}Merging the diabolical points}

\begin{figure}[!b]
\centerline{
\includegraphics[width=0.9\textwidth]{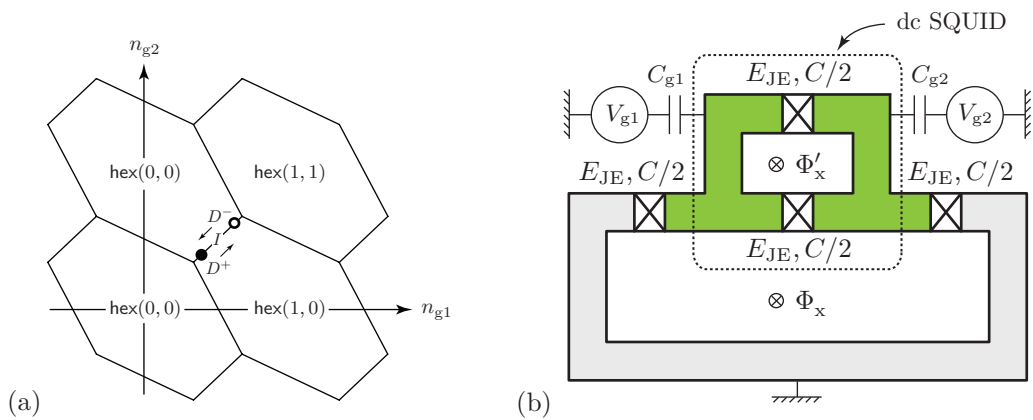}
}
\caption{(a) (Color online) Displacements of two diabolical points $D^+$ and $D^-$ as $E_{\rm J}$ decreases: they merge at their midpoint $I$. (b) The circuit used to merge the points. The central junction of the CPP is replaced by a dc SQUID threaded by a controllable flux $\Phi'_{\rm x}$. It behaves essentially as an effective junction having a capacitance $C$ and a Josephson energy tunable between $0$ and $2\,E_{\rm JE}$.}\label{CPPM}
\end{figure}

As the first approximation in the close vicinity of $T^\pm$, the whole Hilbert space can be reduced to the ground eigenspace of $H_{\rm C}(T^\pm)$. In these three-level models, the positions of $D^\pm$ are easily found. They are located in the plane $\varphi_{\rm x}=\pi$ at
\begin{align*}
{\bf n}_{\rm g}(D^\pm)\approx{\bf n}_{\rm g}(T^\pm)\pm\frac{1}{2E_{\rm JL}E_{\rm J}E_{\rm JR}}\,{\textsf C}\left(\begin{array}{c}E_{\rm JL}^2\big(E_{\rm JR}^2-E_{\rm J}^2\big)\\[1ex]E_{\rm JR}^2\big(E_{\rm JL}^2-E_{\rm J}^2\big)\end{array}\right).
\end{align*}
Suppose that the CPP is mirror symmetric and $E_{\rm J}$ tunable. The above formula illustrates that the displacement of $D^\pm$ conserves the symmetry $[H,\mathcal P]=0$. Furthermore, reducing $E_{\rm J}$ improves $n_{\rm g1}(D^+)=n_{\rm g2}(D^+)$ as much as it reduces $n_{\rm g1}(D^-)=n_{\rm g2}(D^-)$. So, the two diabolical points $D^+$ and $D^-$ shown in figure~\ref{CPPM}(a) are expected to merge symmetrically at their midpoint $I$ located at $(\frac12,\frac12)$ in the ${\bf n}_{\rm g}$-plane.

Let $E_{\rm JE}=E_{\rm JL}=E_{\rm JR}$ be the Josephson energy of the exterior junctions. By the definition of the capacitance unit, their capacitance is $({1-C})/{2}$. Around the point $I$, an approximate Hamiltonian is the restriction of $H$ to the subspace spanned by the basis $\{|0,0\rangle,|1,0\rangle,|0,1\rangle,|1,1\rangle\}$. Up to an unimportant shift of the zero of energies, the truncated Hamiltonian has the form~(\ref{H4}) in this basis, where $X$ and $Y$ are redefined as follows:
\begin{align*}
X=\frac{n_{\rm g2}+n_{\rm g1}-1}{1-C}\, , \qquad Y=\frac{n_{\rm g2}-n_{\rm g1}}{1+3C}\,.
\end{align*}
The other parameters are $\xi={C}/[{(1-C)(1+3C)}]$, $F={E_{\rm JE}}/{2}$ and $G={E_{\rm J}}/{2}$. Within the four-level approximation, if $E_{\rm J}$ is greater than the critical value
\begin{align}
E_{\rm Jc}=\sqrt{2E_{\rm JE}^2+4\xi^2}-2\,\xi=\frac{E_{\rm JE}^2}{2\,\xi}+{\rm O}(E_{\rm JE}^4),\label{EJc}
\end{align}
the degeneracies are located at ${\bf R}^\pm=(\mp X_{\rm d},0,0)$, with
\begin{align*}
X_{\rm d}=\frac12\sqrt{\bigg(1-\frac{E_{\rm Jc}}{E_{\rm J}}\bigg)(E_{\rm J}+4\xi)(E_{\rm J}+E_{\rm Jc}+4\xi)}\,.
\end{align*}

Section~\ref{sec:level2} tells us that the merging of the diabolical points $D^+$ and $D^-$ is possible if $E_{\rm J}$ is adjustable. It is well-known that a tunable effective Josephson coupling can be realized via two junctions in a loop configuration (a dc SQUID). Such a circuit element is {\it de facto} interesting
from the viewpoint of tuning the couplings between superconducting qubits~\cite{Makhlin,Makhlin1}. It has also demonstrated its utility for Cooper pairs pumping in the so-called Cooper pairs sluice~\cite{Pump3}. As shown in figure~\ref{CPPM}~(b), we replace the central junction by a dc SQUID and assume all the junctions of the circuit to be identical. To be consistent with our previous notations, we set $E_{\rm JE}$ to be the Josephson energies and ${C}/{2}$ to be the capacitances of all the junctions. The new central element has a capacitance $C$ and an effective Josephson energy $E_{\rm J}=2\,E_{\rm JE}\big|\cos({\varphi'_{\rm x}}/{2})\big|$, where $\varphi'_{\rm x}=2\pi{\Phi_{\rm x}'}/{\Phi_0}$. The charging and Josephson Hamiltonians still read~(\ref{HC}) and~(\ref{HJ}) after the replacements $\varphi_{\rm x}\to\varphi_{\rm x}+{\varphi_{\rm x}'}/{2}$ and $U(\varphi_{\rm x})\to U(\varphi_{\rm x},\varphi'_{\rm x})$, the exact definition of the last unitary operator being irrelevant for our purpose. We also have $C_{\Sigma}=2\,C$ and $\xi=0.4$.

\begin{figure}[!t]
\centerline{
\includegraphics[width=0.55\textwidth]{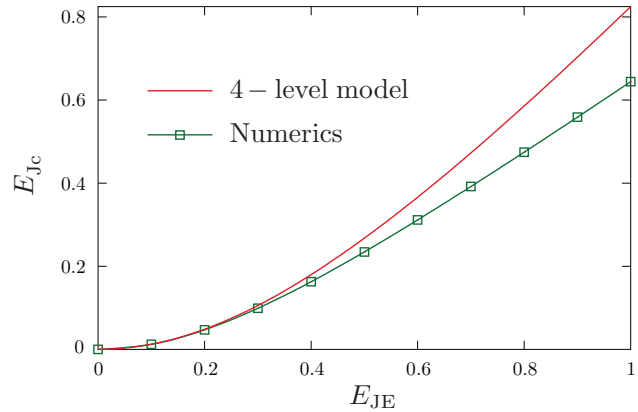}
}
\caption{(Color online) Plots of the critical value $E_{\rm Jc}$ as a function of $E_{\rm JE}$, obtained by the four-level model, on the one hand, and by a numerical treatment, on the other hand. The plot unit is $E_{\rm C}$.}\label{EJcritical}
\end{figure}

The merging is done by tuning the central coupling (through $\varphi'_{\rm x}$) while we use $\varphi_{\rm x}$ to maintain the new half-fluxoid condition $\varphi_{\rm x}+{\varphi_{\rm x}'}/{2}=\pi$. A numerical simulation of the process was made, using the 62 closest states of the point $I$ to define the truncated Hilbert space. The results are in good accordance with the four-level model in the Coulomb blockade regime. For example, in figure~\ref{EJcritical} there is shown  a plot of the critical value $E_{\rm Jc}$ as a function of $E_{\rm J}$: the numerical result coincides with the expression~(\ref{EJc}) in the limit $E_{\rm J}\ll1$.

\section{Conclusion}
We have demonstrated the possibility of merging diabolical points of a superconducting quantum circuit. We have emphasized the role played by the symmetry for that phenomenon. In an experimental perspective, the principal difficulty to overcome is the mirror symmetry since it is impossible to fabricate two identical junctions. This problem can be partially eliminated by using balanced SQUIDs~\cite{Balanced}. The theoretical study was accurately based on a four-level model whose eigenproblem is exactly solvable within the constraint of a degeneracy. For subsequent works, it may serve as a formal model to implement mergings of diabolical points in different contexts, such as quantum circuits or cold atoms. It may also be used to produce non-Abelian holonomies as well as non-Abelian pumpings.

\ukrainianpart

\title{Злиття диявольсьих точок надпровiдного контура}

\author{Р. Леоне,  A. Монжу}

\address{Група статистичної фізики, Інститут ім. Жана Лямура, UMR CNRS 7198, Університет Льорран, Вандувр лє Нансі, Франція}

\makeukrtitle
\begin{abstract}
Представлено перше теоретичне вивчення злиття диявольських точок у застосунку
до надпровiдних контурiв. Спочатку досліджено аналiтично розв’язувану
чотирирiвневу модель, яка може служити теоретичною основою такого явища. В
подальшому ця модель застосовується до контура, який називають помпа
куперiвських пар, з вiдомими диявольськими точками.
\keywords помпа куперiвських пар, диявольські точки, виродженiсть, злиття
\end{abstract}

\lastpage

\begin{thebibliography}{99}

\bibitem{Devoret} Devoret~M.H.,
In: Quantum Fluctuations (Les Houches summer school, Session LXIII, 1995), Reynaud~S., Giacobino~E., Zinn-Justin~J. (Eds.), Elsevier Science B.V., North Holland, Amsterdam, 1997.

\bibitem{Tinkham} Tinkham~M., Introduction to Superconductivity, 2nd Edn., Dover Publications, New York, 2004.

\bibitem{Nakamura} Nakamura~Y., Pashkin~Y.A., Tsai~J.S., Nature, 1999, \textbf{398}, 786; \doi{10.1038/19718}.

\bibitem{Makhlin} {Makhlin~Y., Sch\"on~G., Shnirman~A., Rev. Mod. Phys., 2001, \textbf{73}, 357; \doi{10.1103/RevModPhys.73.357}.}

\bibitem{Martinis} Devoret~M.H., Martinis~J.M., Quantum Inf. Process, 2004, \textbf{3}, 163; \doi{10.1007/s11128-004-3101-5}.

\bibitem{You} You~J.Q., Nori~F., Phys. Today, 2005, \textbf{58}, 42; \doi{10.1063/1.2155757}.

\bibitem{Wendin} Wendin~G., Shumeiko~V.S.,
In: Handbook of Theoretical and Computational Nanotechnology, Vol.~3, Rieth~M., Schommers~W. (Eds.), American Scientific Publishers, Los Angeles, 2006.

\bibitem{Clarke} Clarke~J., Wilhelm~F.K., Nature, 2008, \textbf{453}, 1031; \doi{10.1038/nature07128}.

\bibitem{Blais} Blais~A., Huang R.-S., Wallraff A.,  Girvin S.M.,  Schoelkopf R.J., Phys. Rev. A, 2004, \textbf{69}, 062320; \\ \doi{10.1103/PhysRevA.69.062320}.

\bibitem{Wallraff} Wallraff A.,  Schuster D.I.,  Blais A.,  Frunzio~L.,  Huang~R.-S.,  Majer~J.,  Kumar~S., Girvin~S.M.,  Schoelkopf~R.J., Nature, 2004, \textbf{431}, 162; \doi{10.1038/nature02851}.

\bibitem{Girvin} Schoelkopf~R.J., Girvin~S.M., Nature, 2008, \textbf{451}, 664; \doi{10.1038/451664a}.

\bibitem{Fazio} Faoro~L., Siewert~J., Fazio~R., Phys. Rev. Lett., 2003, \textbf{90}, 028301; \doi{10.1103/PhysRevLett.90.028301}.

\bibitem{Cholascinski} Cholascinski~M., Phys. Rev. B, 2004, \textbf{69}, 134516; \doi{10.1103/PhysRevB.69.134516}.

\bibitem{Pekola2} M\"ott\"onen~M., Pekola J.P., Vartiainen J.J., Brosco~V.,  Hekking~F.W.J., Phys. Rev. B, 2006, \textbf{73}, 214523; \\ \doi{10.1103/PhysRevB.73.214523}.

\bibitem{Leek} Leek~P.J.,  Fink~J.M.,  Blais A.,  Bianchetti~R.,  Goppl~M.,  Gambetta~J.M.,  Schuster~D.I., Frunzio~L.,  Schoelkopf~R.J.,  Wallraff~A., Science, 2007, \textbf{318}, 1889; \doi{10.1126/science.1149858}.

\bibitem{Pekola} M\"ott\"onen~M., Vartiainen~J.J., Pekola~J.P., Phys. Rev. Lett., 2008, \textbf{100}, 177201; \doi{10.1103/PhysRevLett.100.177201}.

\bibitem{Pump1} Geerligs~L.J.,  Verbrugh~S.M.,  Hadley~P.,  Mooij~J.E.,  Pothier~H.,  Lafarge~P.,  Urbina~C., Esteve~D.,  Devoret~M.H., Z. Phys. B: Condens. Matter, 1991, \textbf{85}, 349; \doi{10.1007/BF01307630}.

\bibitem{Pump2} Pekola~J.P.,  Toppari~J.J.,  Aunola~M.,  Savolainen~M.T.,  Averin~D.V., Phys. Rev. B, 1999, \textbf{60}, R9931; \\ \doi{10.1103/PhysRevB.60.R9931}.

\bibitem{Pump3} Niskanen~A.O., Pekola~J.P., Sepp\"a~H., Phys. Rev. Lett., 2003, \textbf{91}, 177003; \doi{10.1103/PhysRevLett.91.177003}.

\bibitem{Pump4} Aunola~M., Toppari~J.J., Phys. Rev. B, 2003, \textbf{68}, 020502; \doi{10.1103/PhysRevB.68.020502}.

\bibitem{Pump5} Vartiainen~J.J., M\"ott\"onen~M., Pekola~J.P., Kemppinen~A., Appl. Phys. Lett., 2007, \textbf{90}, 082102; \doi{10.1063/1.2709967}.

\bibitem{Pump6} Leone~R., L\'evy~L., Phys. Rev. B, 2008, \textbf{77}, 064524; \doi{10.1103/PhysRevB.77.064524}.

\bibitem{Pump7} Brosco~V., Fazio~R., Hekking~F.W.J., Joye~A., Phys. Rev. Lett., 2008, \textbf{100}, 027002; \\ \doi{10.1103/PhysRevLett.100.027002}.

\bibitem{Pump8} Pirkkalainen~J.M., Solinas~P., Pekola~J.P., M\"ott\"onen~M., Phys. Rev. B, 2010, \textbf{81}, 174506; \\ \doi{10.1103/PhysRevB.81.174506}.

\bibitem{Nielsen} Nielsen~M.A., Chuang~I.L., Quantum Computation and Quantum Information, Cambridge University Press, Cambridge, 2000.

\bibitem{MetrologyCPP} Leone~R., L\'evy~L.P., Lafarge~P., Phys. Rev. Lett., 2008, \textbf{100}, 117001; \doi{10.1103/PhysRevLett.100.117001}.

\bibitem{Goryo} Goryo~J., Kohmoto~M., Mod. Phys. Lett. B, 2008, \textbf{22}, 303; \doi{10.1142/S021798490801481X}.

\bibitem{Triangles} Berry~M.V., Wilkinson~M., Proc. R. Soc. London, Ser. A, 1984, \textbf{392}, 15; \doi{10.1098/rspa.1984.0022}.

\bibitem{Montambaux1} Dietl~P., Pi\'echon~F., Montambaux~G., Phys. Rev. Lett., 2008, \textbf{100}, 236405; \doi{10.1103/PhysRevLett.100.236405}.

\bibitem{Montambaux2} Montambaux~G., Pi\'echon~F., Fuchs~J.N., Goerbig~M.O., Phys. Rev. B, 2009, \textbf{80}, 153412; \\ \doi{10.1103/PhysRevB.80.153412}.

\bibitem{Montambaux3} Montambaux~G., Pi\'echon~F., Fuchs~J.N., Goerbig~M.O., Eur. Phys. J. B, 2009, \textbf{72}, 509; \\ \doi{10.1140/epjb/e2009-00383-0}.

\bibitem{Montambaux4} Delplace~P., Montambaux~G., Phys. Rev. B, 2010, \textbf{82}, 035438; \doi{10.1103/PhysRevB.82.035438}.

\bibitem{Optics} Bahat-Treidel~O., Peleg~O., Segev~M., Opt. Lett., 2008, \textbf{33}, 2251; \doi{10.1364/OL.33.002251}.

\bibitem{Pereira} Pereira~V.M., Castro Neto~A.H., Peres~N.M.R., Phys. Rev. B, 2009, \textbf{80}, 045401; \doi{10.1103/PhysRevB.80.045401}.

\bibitem{Asano} Asano~K., Hotta~C., Phys. Rev. B, 2011, \textbf{80}, 245125; \doi{10.1103/PhysRevB.83.245125}.

\bibitem{WVN} Von Neumann~J., Wigner~E.P., Phys. Z., 1929, \textbf{30}, 467.

\bibitem{Simon} Simon~B., Phys. Rev. Lett., 1983, \textbf{51}, 2167; \doi{10.1103/PhysRevLett.51.2167}.

\bibitem{Recati} Recati~A.,  Calarco~T.,  Zanardi~P.,  Cirac~J.I.,  Zoller~P., Phys. Rev. A, 2002, \textbf{66}, 032309; \\ \doi{10.1103/PhysRevA.66.032309}.

\bibitem{LeonePumping} Leone~R., J. Phys. A: Math. Theor., 2011, \textbf{44}, 295301; \doi{10.1088/1751-8113/44/29/295301}.

\bibitem{Makhlin1} Makhlin~Y., Sch\"on~G., Shnirman~A., Nature, 1999, \textbf{386}, 305; \doi{10.1038/18613}.


\bibitem{Balanced} Kemppinen~A.,  Manninen~A.J., M\"ott\"onen~M., Vartiainen~J.J., Peltonen~J.T., Pekola~J.P., Appl. Phys. Lett., 2008, \textbf{92}, 052110; \doi{10.1063/1.2842413}.

\end{thebibliography}
\end{document}